\documentclass[aps,prc,letterpaper,tightenlines,nofootinbib,twocolumn,floatfix]{revtex4}
\usepackage{amssymb,graphicx,amsmath,color,float}
\allowdisplaybreaks

\let\originalleft\left
\let\originalright\right
\renewcommand{\left}{\mathopen{}\mathclose\bgroup\originalleft}
\renewcommand{\right}{\aftergroup\egroup\originalright}

\DeclareSymbolFont{matha}{OML}{txmi}{m}{it}
\DeclareMathSymbol{\varv}{\mathord}{matha}{118}

\ifdefined\myframe
\renewenvironment{myframe}[2]{\section{#1}\begin{frame}{#2}\vspace{-10pt}}{\end{frame}} 
\else
\newenvironment{myframe}[2]{\section{#1}\begin{frame}{#2}\vspace{-10pt}}{\end{frame}}
\fi

\ifdefined\U
\renewcommand{\U}[1]{\underline{#1}}
\else
\newcommand{\U}[1]{\underline{#1}}
\fi

\ifdefined\UU
\renewcommand{\UU}[1]{\underline{\underline{#1}}}
\else
\newcommand{\UU}[1]{\underline{\underline{#1}}}
\fi

\ifdefined\mybullet
\renewcommand{\mybullet}{\vspace{2mm}\\$\bullet$ }
\else
\newcommand{\mybullet}{\vspace{2mm}\\$\bullet$ }
\fi

\ifdefined\mybulletEQ
\renewcommand{\mybulletEQ}[1]{$\bullet$ {\bf #1}\vspace{1mm}\\}
\else
\newcommand{\mybulletEQ}[1]{$\bullet$ {\bf #1}\vspace{1mm}\\}
\fi

\ifdefined\fract
\renewcommand{\fract}[2]{{\textstyle \frac{#1}{#2}}}
\else
\newcommand{\fract}[2]{{\textstyle \frac{#1}{#2}}}
\fi

\ifdefined\rect
\renewcommand{\fract}[1]{{\textstyle \frac{1}{#1}}}
\else
\newcommand{\rect}[1]{{\textstyle \frac{1}{#1}}}
\fi

\ifdefined\fracd
\renewcommand{\fracd}[2]{\frac{\displaystyle{#1}}{\displaystyle{#2}}}
\else
\newcommand{\fracd}[2]{\frac{\displaystyle{#1}}{\displaystyle{#2}}}
\fi

\ifdefined\recd
\renewcommand{\recd}[1]{\frac{\displaystyle 1}{\displaystyle{#1}}}
\else
\newcommand{\recd}[1]{\frac{\displaystyle 1}{\displaystyle{#1}}}
\fi

\ifdefined\pdd
\renewcommand{\pdd}[2]{\frac{\displaystyle{\partial{#1}}}{\displaystyle{\partial{#2}}}}
\else
\newcommand{\pdd}[2]{\frac{\displaystyle{\partial{#1}}}{\displaystyle{\partial{#2}}}}
\fi

\ifdefined\biggg
\renewcommand{\biggg}[1]{\scalebox{1.2}{\Bigg{#1}}}
\else
\newcommand{\biggg}[1]{\scalebox{1.2}{\Bigg{#1}}}
\fi

\ifdefined\Biggg
\renewcommand{\Biggg}[1]{\scalebox{1.4}{\Bigg{#1}}}
\else
\newcommand{\Biggg}[1]{\scalebox{1.4}{\Bigg{#1}}}
\fi

\ifdefined\Re
\renewcommand{\Re}{\operatorname{Re}}
\else
\newcommand{\Re}{\operatorname{Re}}
\fi

\ifdefined\Im
\renewcommand{\Im}{\operatorname{Im}}
\else
\newcommand{\Im}{\operatorname{Im}}
\fi

\ifdefined\Arch
\renewcommand{\Arch}{\operatorname{Ar\,ch}}
\else
\newcommand{\Arch}{\operatorname{Ar\,ch}}
\fi

\ifdefined\Arsh
\renewcommand{\Arsh}{\operatorname{Ar\,sh}}
\else
\newcommand{\Arsh}{\operatorname{Ar\,sh}}
\fi

\ifdefined\Arth
\renewcommand{\Arth}{\operatorname{Arth}}
\else
\newcommand{\Arth}{\operatorname{Arth}}
\fi

\ifdefined\ch
\renewcommand{\ch}{\operatorname{ch}}
\else
\newcommand{\ch}{\operatorname{ch}}
\fi

\ifdefined\sh
\renewcommand{\sh}{\operatorname{sh}}
\else
\newcommand{\sh}{\operatorname{sh}}
\fi

\ifdefined\th
\renewcommand{\th}{\operatorname{th}}
\else
\newcommand{\th}{\operatorname{th}}
\fi

\ifdefined\Ln
\renewcommand{\Ln}{\operatorname{Ln}}
\else
\newcommand{\Ln}{\operatorname{Ln}}
\fi

\ifdefined\tg
\renewcommand{\tg}{\operatorname{tg}}
\else
\newcommand{\tg}{\operatorname{tg}}
\fi

\ifdefined\ctg
\renewcommand{\ctg}{\operatorname{ctg}}
\else
\newcommand{\ctg}{\operatorname{ctg}}
\fi

\ifdefined\intl
\renewcommand{\intl}{\int\limits}
\else
\newcommand{\intl}{\int\limits}
\fi

\ifdefined\ointl
\renewcommand{\ointl}{\oint\limits}
\else
\newcommand{\ointl}{\oint\limits}
\fi

\ifdefined\soint
\renewcommand{\soint}[4]{\hspace{#1}\oint_{#3}^{#4}\hspace{#2}}
\else
\newcommand{\soint}[4]{\hspace{#1}\oint_{#3}^{#4}\hspace{#2}}
\fi

\ifdefined\sointl
\renewcommand{\sointl}[4]{\hspace{#1}\oint\limits_{#3}^{#4}\hspace{#2}}
\else
\newcommand{\sointl}[4]{\hspace{#1}\oint\limits_{#3}^{#4}\hspace{#2}}
\fi

\ifdefined\sint
\renewcommand{\sint}[4]{\hspace{#1}\int_{#3}^{#4}\hspace{#2}}
\else
\newcommand{\sint}[4]{\hspace{#1}\int_{#3}^{#4}\hspace{#2}}
\fi

\ifdefined\sintl
\renewcommand{\sintl}[4]{\hspace{#1}\int\limits_{#3}^{#4}\hspace{#2}}
\else
\newcommand{\sintl}[4]{\hspace{#1}\int\limits_{#3}^{#4}\hspace{#2}}
\fi

\ifdefined\integrated
\renewcommand{\integrated}[3]{\left\{{#1}\right\}\left.\vphantom{#1}\right|_{#2}^{#3}}
\else
\newcommand{\integrated}[3]{\left\{{#1}\right\}\left.\vphantom{#1}\right|_{#2}^{#3}}
\fi

\ifdefined\pd
\renewcommand{\pd}[2]{\frac{\partial{#1}}{\partial{#2}}}
\else
\newcommand{\pd}[2]{\frac{\partial{#1}}{\partial{#2}}}
\fi

\ifdefined\rec
\renewcommand{\rec}[1]{\frac{1}{#1}}
\else
\newcommand{\rec}[1]{\frac{1}{#1}}
\fi

\ifdefined\gvec
\renewcommand{\gvec}[1]{\mbox{\boldmath${#1}$}}
\else
\newcommand{\gvec}[1]{\mbox{\boldmath${#1}$}}
\fi

\ifdefined\cvec
\renewcommand{\cvec}[1]{\mbox{\boldmath${#1}$}}
\else
\newcommand{\cvec}[1]{\mbox{\boldmath${#1}$}}
\fi

\ifdefined\td
\renewcommand{\td}[2]{\frac{d{#1}}{d{#2}}}
\else
\newcommand{\td}[2]{\frac{d{#1}}{d{#2}}}
\fi

\ifdefined\md
\renewcommand{\md}[2]{\frac{\mathrm{d}{#1}}{\mathrm{d}{#2}}}
\else
\newcommand{\md}[2]{\frac{\mathrm{d}{#1}}{\mathrm{d}{#2}}}
\fi

\ifdefined\z
\renewcommand{\z}[1]{\left({#1}\right)}
\else
\newcommand{\z}[1]{\left({#1}\right)}
\fi

\ifdefined\ae
\renewcommand{\ae}[1]{\left|{#1}\right|}
\else
\newcommand{\ae}[1]{\left|{#1}\right|}
\fi

\ifdefined\sz
\renewcommand{\sz}[1]{\left[{#1}\right]}
\else
\newcommand{\sz}[1]{\left[{#1}\right]}
\fi

\ifdefined\kz
\renewcommand{\kz}[1]{\left\{{#1}\right\}}
\else
\newcommand{\kz}[1]{\left\{{#1}\right\}}
\fi

\ifdefined\B
\renewcommand{\B}[1]{\mathbb{#1}}
\else
\newcommand{\B}[1]{\mathbb{#1}}
\fi

\ifdefined\m
\renewcommand{\m}[1]{\mathrm{#1}}
\else
\newcommand{\m}[1]{\mathrm{#1}}
\fi

\ifdefined\tn
\renewcommand{\tn}[1]{\textnormal{#1}}
\else
\newcommand{\tn}[1]{\textnormal{#1}}
\fi

\ifdefined\o
\renewcommand{\o}[1]{\operatorname{#1}}
\else
\newcommand{\o}[1]{\operatorname{#1}}
\fi

\ifdefined\c
\renewcommand{\c}[1]{\mathcal{#1}}
\else
\newcommand{\c}[1]{\mathcal{#1}}
\fi

\ifdefined\f
\renewcommand{\f}[1]{\mathfrak{#1}}
\else
\newcommand{\f}[1]{\mathfrak{#1}}
\fi

\ifdefined\v
\renewcommand{\v}[1]{\mathbf{#1}}
\else
\newcommand{\v}[1]{\mathbf{#1}}
\fi

\ifdefined\bs
\renewcommand{\bs}[1]{\boldsymbol{#1}}
\else
\newcommand{\bs}[1]{\boldsymbol{#1}}
\fi

\ifdefined\Eq
\renewcommand{\Eq}[1]{Eq.~(\ref{#1})}
\else
\newcommand{\Eq}[1]{Eq.~(\ref{#1})}
\fi

\ifdefined\Eqs
\renewcommand{\Eqs}[2]{Eqs.~(\ref{#1}) and (\ref{#2})}
\else
\newcommand{\Eqs}[2]{Eqs.~(\ref{#1}) and (\ref{#2})}
\fi

\ifdefined\a
\renewcommand{\a}[1]{\aref({#1})}
\else
\newcommand{\a}[1]{\aref({#1})}
\fi

\ifdefined\A
\renewcommand{\A}[1]{\Aref({#1})}
\else
\newcommand{\A}[1]{\Aref({#1})}
\fi

\ifdefined\r

\renewcommand{\r}[1]{(\ref{#1})}
\else
\newcommand{\r}[1]{(\ref{#1})}
\fi

\ifdefined\comm
\renewcommand{\comm}[2]{\left[{#1},{#2}\right]}
\else
\newcommand{\comm}[2]{\left[{#1},{#2}\right]}
\fi

\ifdefined\follows
\renewcommand{\follows}{\quad\Rightarrow\quad}
\else
\newcommand{\follows}{\quad\Rightarrow\quad}
\fi

\ifdefined\Follows
\renewcommand{\Follows}{\qquad\Rightarrow\qquad}
\else
\newcommand{\Follows}{\qquad\Rightarrow\qquad}
\fi

\ifdefined\followse
\renewcommand{\followse}{\quad\Rightarrow}
\else
\newcommand{\followse}{\quad\Rightarrow}
\fi

\ifdefined\bfollows
\renewcommand{\bfollows}{\Rightarrow\quad}
\else
\newcommand{\bfollows}{\Rightarrow\quad}
\fi

\ifdefined\equivalent
\renewcommand{\equivalent}{\quad\Leftrightarrow\quad}
\else
\newcommand{\equivalent}{\quad\Leftrightarrow\quad}
\fi

\ifdefined\Equivalent
\renewcommand{\Equivalent}{\qquad\Leftrightarrow\qquad}
\else
\newcommand{\Equivalent}{\qquad\Leftrightarrow\qquad}
\fi

\ifdefined\obs
\renewcommand{\obs}[1]{\left\langle{#1}\right\rangle}
\else
\newcommand{\obs}[1]{\left\langle{#1}\right\rangle}
\fi

\ifdefined\ket
\renewcommand{\ket}[1]{\left|{#1}\right\rangle}
\else
\newcommand{\ket}[1]{\left|{#1}\right\rangle}
\fi

\ifdefined\bra
\renewcommand{\bra}[1]{\left\langle{#1}\right|}
\else
\newcommand{\bra}[1]{\left\langle{#1}\right|}
\fi

\ifdefined\braket
\renewcommand{\braket}[2]{\left<#1\vphantom{#2}\right|\left.#2\vphantom{#1}\right>}
\else
\newcommand{\braket}[2]{\left<#1\vphantom{#2}\right|\left.#2\vphantom{#1}\right>}
\fi

\ifdefined\ketbra
\renewcommand{\ketbra}[2]{\left|#1\vphantom{#2}\right>\left<#2\vphantom{#1}\right|}
\else
\newcommand{\ketbra}[2]{\left|#1\vphantom{#2}\right>\left<#2\vphantom{#1}\right|}
\fi

\ifdefined\scalprod
\renewcommand{\scalprod}[2]{\left(#1\vphantom{#2}\right|\left.#2\vphantom{#1}\right)}
\else
\newcommand{\scalprod}[2]{\left(#1\vphantom{#2}\right|\left.#2\vphantom{#1}\right)}
\fi

\ifdefined\fixmatrix
\renewcommand{\fixmatrix}[2]{\left(\begin{array}{*{9}{@{}>{\centering\arraybackslash $}m{#1}<{$ }@{}}}#2\end{array}\right)}
\else
\newcommand{\fixmatrix}[2]{\left(\begin{array}{*{9}{@{}>{\centering\arraybackslash $}m{#1}<{$ }@{}}}#2\end{array}\right)}
\fi

\ifdefined\fixgausselim
\renewcommand{\fixgausselim}[4]{\left(\hspace{-1mm}\begin{array}{*{9}{@{}>{\centering\arraybackslash $}m{#1}<{$ }@{}}}#3\end{array}\vphantom{\begin{array}{*{100}c}#4\end{array}}\hspace{-1mm}\right|\hspace{-1mm}\left.\begin{array}{*{9}{@{}>{\centering\arraybackslash $}m{#2}<{$ }@{}}}#4\end{array}\vphantom{\begin{array}{*{100}c}#3\end{array}}\right)}
\else
\newcommand{\fixgausselim}[4]{\left(\hspace{-1mm}\begin{array}{*{9}{@{}>{\centering\arraybackslash $}m{#1}<{$ }@{}}}#3\end{array}\vphantom{\begin{array}{*{100}c}#4\end{array}}\hspace{-1mm}\right|\hspace{-1mm}\left.\begin{array}{*{9}{@{}>{\centering\arraybackslash $}m{#2}<{$ }@{}}}#4\end{array}\vphantom{\begin{array}{*{100}c}#3\end{array}}\right)}
\fi

\ifdefined\gausselim
\renewcommand{\gausselim}[2]{\left(\begin{matrix}#1\end{matrix}\vphantom{\begin{matrix}#2\end{matrix}}\hspace{1mm}\right|\left.\begin{matrix}#2\end{matrix}\vphantom{\begin{matrix}#1\end{matrix}}\right)}
\else
\newcommand{\gausselim}[2]{\left(\begin{matrix}#1\end{matrix}\vphantom{\begin{matrix}#2\end{matrix}}\hspace{1mm}\right|\left.\begin{matrix}#2\end{matrix}\vphantom{\begin{matrix}#1\end{matrix}}\right)}
\fi

\ifdefined\matrixel
\renewcommand{\matrixel}[3]{\left<#1\vphantom{#2#3}\right|#2\left|#3\vphantom{#1#2}\right>} 
\else
\newcommand{\matrixel}[3]{\left<#1\vphantom{#2#3}\right|#2\left|#3\vphantom{#1#2}\right>} 
\fi

\ifdefined\contravcov
\renewcommand{\contravcov}[3]{{{#1}^{#2}_{}}_{#3}}
\else
\newcommand{\contravcov}[3]{{{#1}^{#2}_{}}_{#3}}
\fi

\ifdefined\covcontrav
\renewcommand{\covcontrav}[3]{{{#1}_{#2}^{}}^{#3}}
\else
\newcommand{\covcontrav}[3]{{{#1}_{#2}^{}}^{#3}}
\fi

\ifdefined\am
\renewcommand{\am}{{\hat{a}^{\vphantom\dagger}}}
\else
\newcommand{\am}{{\hat{a}^{\vphantom\dagger}}}
\fi

\ifdefined\ap
\renewcommand{\ap}{{\hat{a}^\dagger}}
\else
\newcommand{\ap}{{\hat{a}^\dagger}}
\fi

\ifdefined\bm
\renewcommand{\bm}{{\hat{b}^{\vphantom\dagger}}}
\else
\newcommand{\bm}{{\hat{b}^{\vphantom\dagger}}}
\fi

\ifdefined\bp
\renewcommand{\bp}{{\hat{b}^\dagger}}
\else
\newcommand{\bp}{{\hat{b}^\dagger}}
\fi

\ifdefined\arctg
\renewcommand{\arctg}{\operatorname{arctg}}
\else
\newcommand{\arctg}{\operatorname{arctg}}
\fi

\begin{document} 
\title{Coulomb and strong interactions in the final state of HBT correlations\\for L\'evy type source functions}
\author{D.~Kincses$^{1}$, M.~I.~Nagy$^{1}$ and M. Csan\'ad$^{1}$\\\textit{correspondence: kincses@ttk.elte.hu}\vspace{-5pt}}
\affiliation{$^1$ E{\"o}tv{\"o}s Lor{\'a}nd University, H-1117 Budapest, P{\'a}zm{\'a}ny P. s. 1/A, Hungary}

\begin{abstract}
We present detailed calculations about the expected shape of two-pion Bose-Einstein (or HBT) correlations in high energy heavy ion collisions that include a realistic treatment of final state Coulomb interaction as well as strong interactions (dominated by s-wave scattering). We assume L\'evy type source functions, a generalization that goes beyond the Gaussian approximation. Various recent experimental results found the use of such source types necessary to properly describe the shape of the measured correlation functions. We find that strong final state interactions may play an important role in the shape of the two-pion correlation functions, especially if one considers source parameters beyond the Gaussian HBT radii. Precise experimental determination of these source parameters (such as L\'evy stability exponent, correlation strength, etc.) seem to require the inclusion of the treatment of strong interaction not just for heavier particles (e.g. protons, lambdas) but also in case of two-pion measurements.
\end{abstract}

\maketitle

\section{Introduction}

Heavy ion physics strives to understand the properties of strongly interacting matter produced in high energy nuclear collisions. One of the key observables suited for the experimental investigation of the space-time geometry of such collision events is the femtoscopic correlation of produced particles (called Bose-Einstein correlations in case of identical bosons). Since the discovery of quantum statistical correlations of pions produced in high energy reactions~\cite{Goldhaber:1959mj,Goldhaber:1960sf}, more and more experimental data led to a refined understanding of the connection between such correlations and the actual source dynamics, as well as an increased expectation on phenomenological models to reproduce the observations. In conjunction with the discovery of the strongly interacting Quark-Gluon Plasma (sQGP) by the experiments at the Relativistic Heavy Ion Collider~\cite{Adcox:2004mh,Adams:2005dq,Arsene:2004fa,Back:2004je} a renewed interest arose in the investigation of femtoscopic correlations. For a review of such measurements and connected phenomenological studies, see e.g. Refs.~\cite{Lisa:2005dd,Csorgo:1999sj}.

In heavy ion physics, for many years the usual assumption for the source shape was Gaussian. This was corroborated by phenomenological studies such as hydrodynamical model calculations (see e.g. Refs.~\cite{Csorgo:1994fg,Akkelin:1995gh}). Recent results showed that to achieve a statistically acceptable description of the measured correlation functions, one must go beyond this simple picture. The application of the source imaging technique discussed in Ref.~\cite{Brown:1997ku} to 
correlation functions measured in high energy heavy ion collisions led to one of the first signs of non-Gaussian behavior in such reactions~\cite{Adler:2006as}; it was found that the two-pion source function indeed exhibits a power-law behavior. Since then a lot of experimental as well as theoretical work has been done in this direction. Recent results by the PHENIX experiment~\cite{Adare:2017vig} showed that by utilizing L\'evy type sources one can provide an acceptable description of the measured correlations. These type of source functions are expected to emerge from a scenario called anomalous diffusion~\cite{Csanad:2007fr}, but there are other possible competing explanations such as jet fragmentation~\cite{Csorgo:2004sr} or critical behavior~\cite{Csorgo:2005it}.

When one tries to extract information about the source through the analysis of femtoscopic correlations, it is of utmost importance to properly take into account final state interactions (FSI). The shapes of the experimentally measured correlation functions are significantly affected by these interactions (such as Coulomb repulsion and also strong interaction), and taking them into account in the theoretical framework is crucial. The effect of the Coulomb interaction and the methods to properly include it in the description of the correlation functions have been widely studied before, for details see e.g. Refs.~\cite{Sinyukov:1998fc,Lednicky:2005tb,Alt:1998nr}. However, final state strong interaction between like-sign pions is generally thought to have a small effect~\cite{Pratt:1990zq}, so in most experimental analyses it is neglected. In this paper we present a detailed calculation of the shape of two-pion HBT correlation functions with the assumption of L\'evy stable source functions taking into account Coulomb and strong final state interactions.

The structure of the paper is as follows: in Section \ref{s:femto}. we discuss the basic definitions and properties of the femtoscopic correlations with special emphasis on the choice of the source function. In Section \ref{s:fsi}. we investigate the effect of final state interactions on the pair wave function, and subsequently on the correlation function. In Section~\ref{s:results}. we present results of a numerical calculation of the correlation function and investigate the differences between using only Coulomb or both Coulomb and strong interactions. Finally, in Section \ref{s:summary}. we conclude and summarize our findings.

\section{Femtoscopic correlations}
\label{s:femto}

In this section we discuss the basic definitions and properties of femtoscopic correlations, with special emphasis on the shape of the source function.

\subsection{Basic definitions}
\label{ss:def}
The general definition of the two-particle correlation function as a function of the single particle four-momenta is the following:
\begin{equation}
C_2(p_1,p_2) = \frac{N_2(p_1,p_2)}{N_1(p_1)N_1(p_2)},
\label{eq:c2def}
\end{equation}
where $N_1(p_1), N_1(p_2)$ and $N_2(p_1,p_2)$ are the one- and two-particle invariant momentum distributions. The pair momentum distribution can be calculated from the $S(x,p)$ source distribution and the $\Psi^{(2)}_{p_1,p_2}(x_1,x_2)$ symmetrized pair wave function:
\begin{equation}
N_2(p_1,p_2){=}\sint{-1mm}{-3mm}\;\;\; d^4x_1d^4x_2S(x_1,p_1)S(x_2,p_2)\big|\Psi^{(2)}_{p_1,p_2}(x_1,x_2)\big|^2.
\end{equation}
Using the pair source $D(r,K)$, defined as
\begin{equation}
D(r,K) = \int S(\rho + r/2,K)S(\rho-r/2,K)d^4\rho,
\end{equation}
equation (\ref{eq:c2def}) can be reinterpreted as
\begin{equation}
C_2(k,K) = \int d^4r D(r,K)\big|\Psi^{(2)}_{k}(r)\big|^2.
\end{equation}
This way, instead of the single-particle variables $p_1,p_2,x_1,x_2$ one can use the following pair variables: the pair separation four-vector $r$, the pair center of mass four-vector $\rho$, the relative momentum $k = (p_1 - p_2)/2$, and the average momentum $K = (p_1 + p_2)/2$. Since the Lorentz-product of the $k$ and $K$ four-vectors are zero, one may transform the $k$ dependent correlation function to depend on the three-vector component $\bs k$ only. Furthermore, if the energy of the particles contributing to the correlation function are similar, then $K$ is approximately on shell, so the correlation function can be measured as a function of $\bs k$ and $\bs K$.

At this point it is also useful to introduce the core-halo picture, in which the particle emitting source has two components: a hydrodynamically behaving fireball-like core which contains particles created directly from the freeze-out (or from decays of short-lived resonances), and a surrounding halo which contains particles that are the decay products of long-lived resonances (such as $\eta, \eta', K_S^0, \omega$). This picture is particularly important for pions, but the general structure of the model may be relevant for other mesons as well. If one assumes that the single-particle source has two components ($S=S_{core}+S_{halo}$), it follows that the pair source $D$ will have three - a core-core, a core-halo, and a halo-halo component:
\begin{equation}
D = D_{(c,c)}+D_{(c,h)}+D_{(h,h)}.
\end{equation}
Experimentally however, only the core-core part is relevant, the width of the Fourier transform of the other two is below the minimal resolvable momentum difference. Introducing the correlation strength parameter $\lambda$ and coupling the core-halo model with the Bowler-Sinyukov procedure the correlation function can be written as 
\begin{equation}
C_2(\bs k,\bs K) = 1-\lambda+\lambda\int d^3\bs r D_{(c,c)}(\bs r,\bs K)\big|\Psi^{(2)}_{\bs k}(\bs r)\big|^2.
\label{eq:c2}
\end{equation}
More details about the core-halo model and the importance of the $\lambda$ correlation strength parameter can be found e.g. in Ref.~\cite{Adare:2017vig} .

To calculate the shape of the $C_2(\bs k,\bs K)$ two-particle correlation function, one needs an assumption on the shape of the pair source $D_{(c,c)}(\bs r,\boldsymbol K)$, and a proper description of the effect of final state interactions enclosed in the $\Psi^{(2)}_{\bs k}(\bs r)$ pair wave function. In the following, in Section \ref{ss:levysources}. we discuss the details of L\'evy type source functions, and in Section \ref{ss:wavefunc}. we proceed by discussing the calculation of $\Psi^{(2)}_{\bs k}(\bs r)$ with the Coulomb and strong final state interactions included. Finally, in Section \ref{ss:corrcalc}. we combine the previous calculations to derive the shape of the correlation function.

\subsection{L\'evy-stable source functions}
\label{ss:levysources}

Stable distributions are of utmost importance when studying the limiting distributions of random variables based on a sum of elementary processes. It is well known, that in case of one dimensional random variables, the stable distributions can be given through the following formula:
\begin{equation}
f(x; \alpha, \beta, R, \mu) =\frac{1}{2\pi} \int_{-\infty}^{\infty}\varphi(q; \alpha, \beta, R, \mu)e^{iqx}dq,
\label{eq:fx}
\end{equation}
where the characteristic function is given as:
\begin{align}
\varphi(q; \alpha, \beta, R, \mu) &= \exp\left(iq\mu-|qR|^\alpha(1-i\beta\textnormal{sgn}(q)\Phi)\right), \\\nonumber\textnormal{where } \Phi &= \left\{\begin{array}{l}\tan(\frac{\pi\alpha}{2}), \alpha \neq 1,\\-\frac{2}{\pi}\log|q|, \alpha = 1.\end{array}\right. 
\label{eq:phi}
\end{align}
In our case, the symmetric, centered ($\beta=0$, $\mu=0$) stable distributions may play a role of the source distribution, if that results from a statistical process. In multiple dimensions, the situation is far less clear. It is however known that the following distribution in $N$ dimensions is stable~\cite{oroszref}:
\begin{equation}
\c L(\bs r; \alpha, R) =\frac{1}{(2\pi)^3} \int d^3\bs q e^{i\bs q\bs r}e^{-\rec{2}|\bs q\v R\bs q|^{\alpha/2}},
\end{equation}
from which in case of spherical symmetry ($R_{ij} = R^2 \delta_{ij}$), we obtain
\begin{equation}
\c L(\bs r; \alpha, R) =\frac{1}{(2\pi)^3} \int d^3\bs q e^{i\bs q\bs r}e^{-\rec{2}|\bs qR|^\alpha}.
\label{eq:fx}
\end{equation}

\begin{figure}
\includegraphics[width=0.5\textwidth]{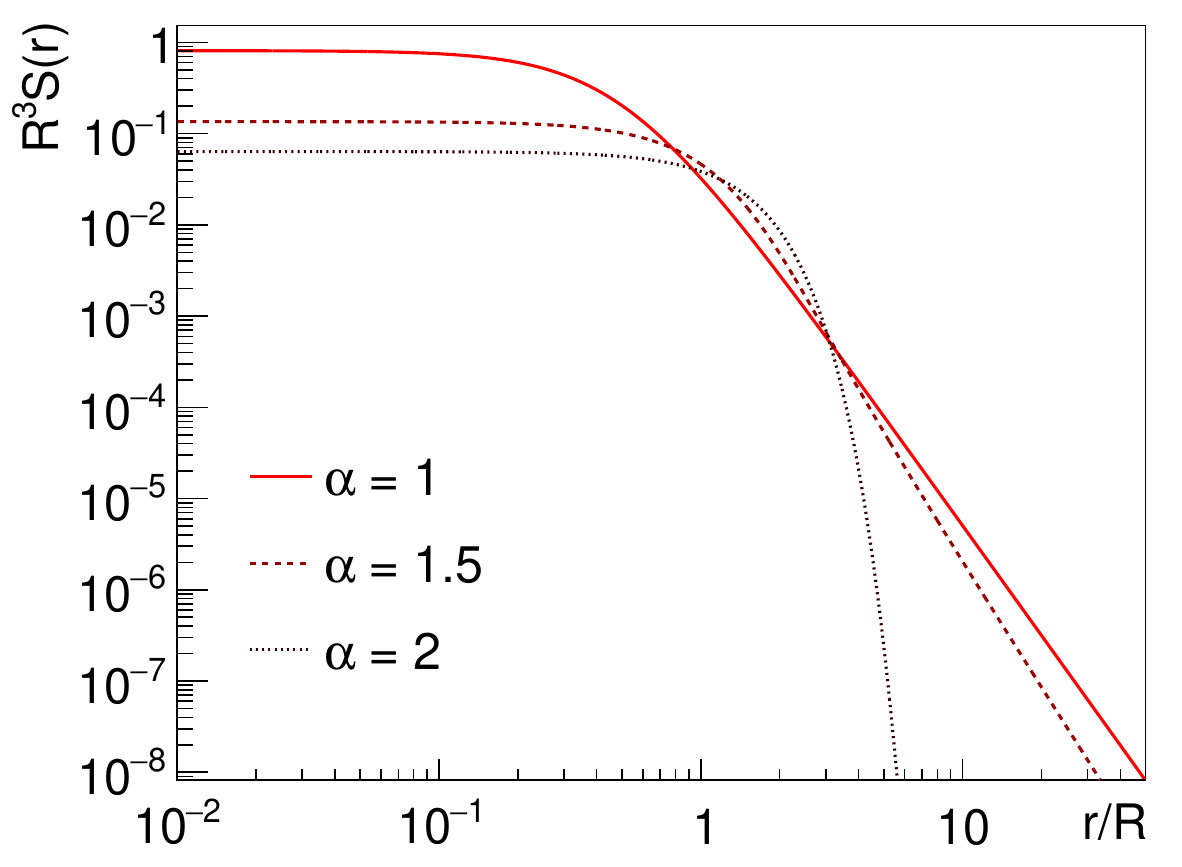}
\caption{L\'evy-stable source distributions with $S(r) = \c L(|\bs r|; \alpha,R)$ for $\alpha =$ 1, 1.5 and 2. The dependence on $R$ is scaled out.}
\label{f:levyexample}
\end{figure}

The two main parameters of such distributions are the index of stability, $\alpha$, and the scale parameter, $R$. In case of $\alpha < 2$ the distribution exhibits a power-law behavior, while the $\alpha = 2$ case corresponds to the Gaussian distribution. The most important property of this distribution is that any moment greater than $\alpha$ is not defined and it retains the same $\alpha$ under convolution of random variables. From the latter it is apparent that if the single particle source $S_{core}(\bs r)$ is a L\'evy-stable distribution, then the pair-source $D_{(c,c)}(\bs r)$ also has a L\'evy shape with the same index of stability $\alpha$:
\begin{equation}
S_{core}(\bs r) = \c L(\bs r; \alpha, R) \Rightarrow D_{(c,c)}(\bs r) = \c L(\bs r; \alpha, 2^{1/\alpha}R)
\end{equation}
An illustration of the shape of such distributions can be seen on Fig.~\ref{f:levyexample}. The average momentum dependence appears through the two parameters of $D_{(c,c)}(\bs r)$:
\begin{equation}
D_{(c,c)}(\bs r, \bs K) = \c L(\bs r; \alpha(\bs K), 2^{1/\alpha(\bs K)}R(\bs K)).
\label{eq:dcc}
\end{equation}

The dependence of the L\'evy source parameters on the pair average momentum $K$ is non-trivial, and is often the subject of the experimental investigations.

\section{Final state interactions}
\label{s:fsi}

To make the paper as self-contained as possible, in this section we review the methodology of the calculation of a correlation function that includes the effect of the final state Coulomb and strong interactions. In doing so, we closely follow along the lines of Ref.~\cite{Lednicky:2005tb}.

\subsection{The pair wave function}
\label{ss:wavefunc}

Firstly let us introduce the Sommerfeld parameter $\eta$ that appears frequently during calculations concerning the quantum mechanical Coulomb problem:
\begin{align}
\eta \equiv \frac{q_e^2}{4\pi\varepsilon_0}\frac{\mu}{\hbar^2k},\quad \mu = \frac{m_1m_2}{m_1{+}m_2}.
\end{align}
Here $\mu$ is the reduced mass of the particle pair. Note that one often uses the fine structure constant $\alpha \equiv \frac{q_e^2}{4\pi\varepsilon_0}\rec{\hbar c} \approx \rec{137}$ in the definition of $\eta$, with which it could be written as $\eta = \frac{\mu\alpha}{\hbar ck}$; nevertheless, we avoid this in this paper because we denote the L\'evy index also by $\alpha$, as indicated in the previous section.

A normalization constant $\c N$ appears in many contexts in the Coulomb wave function. Its definition is
\begin{align}
\c N = e^{-\pi\eta/2}\Gamma(1{+}i\eta),
\end{align}
and its modulus square, which is called the Gamow factor, can be calculated with elementary functions (owing to the well known step and reflection properties of the gamma function) as 
\begin{align}
|\c N|^2 &= \frac{2\pi\eta}{e^{2\pi\eta}{-}1} .
\end{align}
The Schr\"odinger equation in a repulsive Coulomb potential can be written as
\begin{align}
\label{e:Schreq}
\triangle\psi_{\bs k}(\bs r) - \frac{2\eta k}{r}\psi_{\bs k}(\bs r) = k^2\psi_{\bs k}(\bs r).
\end{align}
For the treatment of the final state interactions, one has to utilize the scattering wave solutions whose asymptotic form is a plane wave plus a spherical wave. Such solutions for the Coulomb potential are well known: 
\begin{align}
\psi_{\bs k}^{(+)}(\bs r) &= \c Ne^{i\bs k\bs r}\v F(-i\eta,1,i(kr{-}\bs k\bs r))=\nonumber\\&= \c Ne^{ikr}\v F(1{+}i\eta,1,-i(kr{-}\bs k\bs r)),\\
\psi_{\bs k}^{(-)}(\bs r) &= \c N^*e^{i\bs k\bs r}\v F(i\eta,1,-i(kr{+}\bs k\bs r))=\nonumber\\&= \c N^*e^{-ikr}\v F(1{-}i\eta,1,i(kr{+}\bs k\bs r)).
\end{align}
Here $\v F(a,b,z)$ is the (renormalized) confluent hypergeometric function (Kummer's function); its definition and some basic properties are recited in Appendix~\ref{s:app:formulas}. (A well-known property shows that the two forms of each functions introduced here are indeed equal.)

The connection between these wave functions is
\begin{align}
\psi_{\bs k}^{(+)}(\bs r) = \big(\psi_{-\bs k}^{(-)}(\bs r)\big)^*.
\end{align}
From the asymptotic expression of the confluent hypergeometric function one can verify that the asymptotic form of these wave functions is
\begin{align}
\psi_{\bs k}^{(+)}(\bs r) \approx\ &e^{i\bs k\bs r}e^{i\eta\log(kr-\bs k\bs r)} +\nonumber\\ &+f_c(\vartheta)\frac{e^{ikr}}{r}e^{-i\eta\log(kr-\bs k\bs r)},\\
\psi_{\bs k}^{(-)}(\bs r) \approx\ &e^{i\bs k\bs r}e^{-i\eta\log(kr+\bs k\bs r)} +\nonumber\\ &+ f^*_c(\vartheta)\frac{e^{-ikr}}{r}e^{i\eta\log(kr+\bs k\bs r)}.
\end{align}
Here the notation $f_c(\vartheta)$ stands for the Coulomb scattering amplitude, which is defined as
\begin{align}
f_c(\vartheta) = - \frac{\eta}{2k}\rec{\sin^2\frac\vartheta2}\frac{\Gamma(1{+}i\eta)}{\Gamma(1{-}i\eta)} .
\end{align}
One indeed sees that asymptotically the $\psi_{\bs k}^{(+)}(\bs r)$ and the $\psi_{\bs k}^{(-)}(\bs r)$ wave functions contain a plane wave plus an outgoing or an incoming spherical wave, respectively. (There are logarithmic factors stemming from the long range nature of the Coulomb interaction that distort both of them; these factors do not influence the physical meaning of the wave functions.) The $\psi_{\bs k}^{(+)}(\bs r)$ and the $\psi_{\bs k}^{(-)}(\bs r)$ functions are called {\it in} and {\it out} scattering states, respectively.\footnote{It is a known fact that when calculating transition matrix elements, one has to utilize the $\psi_{\bs k}^{(-)}(\bs r)$ state (the {\it out} state) for the wave function of the final state; this might seem somewhat counter-intuitive, since this function contains an incoming spherical wave. Similarly, one has to use $\psi_{\bs k}^{(+)}(\bs r)$ for the initial state. See e.g. Ref.~\cite{BetheLow} for some details.}

The scattering states written up here can be expanded in terms of energy eigenstates which are also angular momentum eigenstates. For given $l$ and $m$ angular momentum quantum numbers, one has two linearly independent angular momentum eigenstate solutions of the (\ref{e:Schreq}) Schr\"odinger equation: their angle dependence is that of the $Y_{lm}(\vartheta,\varphi)$ spherical harmonic function, and their radial parts are called {\it regular} and {\it singular} Coulomb waves, respectively. We denote them here by $\c F_{k,l}(r)$ and $\c G_{k,l}(r)$ (as they depend on the $k$ wave number magnitude and the $l$ total angular momentum quantum number but not on the magnetic quantum number $m$); their expression is
\begin{align}
\c F_{k,l}(r) &= e^{\pi\eta/2}(-1)^{l+1}4k(2kr)^l\times\nonumber\\\times&\f R\Big\{e^{ikr+i\delta^c_{k,l}}{\times}U\big(l{+}1{+}i\eta,2l{+}2,-2ikr\big)\Big\},\\
\c G_{k,l}(r) &= -e^{\pi\eta/2}(-1)^{l+1}4k(2kr)^l\times\nonumber\\\times&\f I\Big\{e^{ikr+i\delta^c_{k,l}}{\times}U\big(l{+}1{+}i\eta,2l{+}2,-2ikr\big)\Big\},
\end{align}
where the so-called Tricomi's function, $U(a,b,z)$ is another solution of the confluent hypergeometric equation (see Appendix~\ref{s:app:formulas} for some details). They are chosen for the set of linearly independent solutions because $\c F_{k,l}$ is finite at the $r{=}0$ origin, and their asymptotic form is quite simple and straightforward: for $r\to\infty$ we have
\begin{align}
\c F_{k,l}(r) &\approx\frac 2r\sin\z{kr{-}\frac{l\pi}{2}{+}\delta_{k,l}^c-\eta\log(2kr)},\\
\c G_{k,l}(r) &\approx\frac 2r\cos\z{kr{-}\frac{l\pi}{2}{+}\delta_{k,l}^c-\eta\log(2kr)},
\end{align}
where the so-called Coulomb phase shift $\delta_{k,l}^c$ is defined as
\begin{align}
\delta^c_{k,l}\equiv\arg\Gamma(l{+}1{+}i\eta).
\end{align}
One can also take a linear combination of these two functions whose asymptotic form contains an additional arbitrary $\Delta_{k,l}$ phase shift.
\begin{align}
\c M_{k,l}(r) := \cos\Delta_{k,l}\cdot\c F_{k,l}(r){+}\sin\Delta_{k,l}\cdot\c G_{k,l}(r),
\end{align}
whose asymptotic form duly is
\begin{align}
\c M_{k,l}(r)\approx\frac 2r\sin\z{kr{-}\frac{l\pi}{2}{+}\Delta_{k,l}{+}\delta_{k,l}^c{-}\eta\log(2kr)}.
\end{align}
The above scattering-like solutions of the Schr\"odinger equation can be expanded in partial waves as
\begin{align}
\label{e:partwave}
\psi_{\bs k}^{(-)}(\bs r) = \sum_{l=0}^\infty\frac{2l{+}1}{2k}(-i)^lP_l(\cos\vartheta)e^{-i\delta_l^c}\c F_{k,l}(r).
\end{align}
Owing to the short range of strong interaction, we can treat its effect by introducing the $\Delta^s_{k,0}$ s-wave ,,strong'' phase shift, and modifying the s-wave component of the exact Coulomb wave function to a s-wave which contains this additional phase shift (see more details in e.g. Ref~\cite{LandauIII}). This is done by replacing the $\c F_{k,0}$ function in the $l{=}0$ term in the expansion (\ref{e:partwave}) with the above defined $\c M_{k,0}^s(r)$ function which contains the additional $\Delta^s_{k,0}$ phase shift: 
\begin{align}
\psi_{\bs k}^{(-)}(\bs r) \quad\to\quad \Psi_{\bs k}^{\m{cs}}(\bs r),
\end{align}
so the wave function incorporating the Coulomb and strong interaction effects, $\Psi_{\bs k}^{\m{cs}}(\bs r)$, becomes
\begin{align}
&\Psi_{\bs k}^{\m{cs}}(\bs r) = \psi_{\bs k}^{(-)}(\bs r)-\frac{e^{-i\delta_{k,0}^c}}{2k}\c F_{k,0}(r) +\nonumber\\ &\qquad\qquad+\frac{e^{-i\delta_{k,0}^c}}{2k}e^{-i\Delta^s_{k,0}}\c M_{k,0}^s(r)= \nonumber\\
&{=}\ \psi_{\bs k}^{(-)}(\bs r){-}\frac{i}{2k}e^{-i(\delta_{k,0}^c+\Delta^s_{k,0})}\sin\Delta^s_{k,0}\big(\c F_{k,0}{+}i\c G_{k,0}\big).
\end{align}
Substituting the formulas for the respective wave functions encountered here, we get
\begin{align}
&\Psi_{\bs k}^{\m{cs}}(\bs r) = e^{-ikr}\Big\{\c N^*\v F\big(1{-}i\eta,1,i(kr{+}\bs k\bs r)\big) + \nonumber\\&{+}2i\sin\Delta_{k,0}^se^{-i\Delta_{k,0}^s}e^{\pi\eta/2}e^{-2i\delta_{k,0}^c}U\big(1{-}i\eta,2,2ikr\big)\Big\} .
\end{align}
For identical bosonic particles (e.g. pions) one needs the symmetrized two-particle wave function:
\begin{alignat}{2}
\label{eq:psi}
&\Psi^{(2)}_{\bs k}(\bs r) &&{:=} \rec{\sqrt2}\big(\Psi_{\bs k}^{\m{cs}}(\bs r)+\Psi_{\bs k}^{\m{cs}}(-\bs r)\big) =\nonumber\\&=\frac{e^{-ikr}}{\sqrt2}&&\Big\{\c N^*\v F\big(1{-}i\eta,1,i(kr{+}\bs k\bs r)\big) +\nonumber\hfill\\&&&{+}\c N^*\v F\big(1{-}i\eta,1,i(kr{-}\bs k\bs r)\big) +\nonumber\\&{+}4i\sin\Delta&&_{k,0}^se^{-i\Delta_{k,0}^s}e^{\pi\eta/2}e^{-2i\delta_{k,0}^c}U\big(1{-}i\eta,2,2ikr\big)\Big\} .
\end{alignat}
Finally, one needs to calculate the modulus square of the wave function. The $[\bs r\to-\bs r]$ term within the braces in the following expression represents terms similar to the ones that stand before it, just with a mirrored $\bs r$:\pagebreak
\begin{widetext}
\begin{align}
\label{eq:psiabs}
\big|\Psi^{(2)}_{\bs k}(\bs r)\big|^2 =&\Bigg\{\frac{|\c N|^2}{2}\big|\v F\big(1{-}i\eta,1,i(kr{+}\bs k \bs r)\big)\big|^2+\frac{|\c N|^2}{2}\v F\big(1{+}i\eta,1,-i(kr{+}\bs k \bs r)\big)\v F\big(1{-}i\eta,1,i(kr{-}\bs k \bs r)\big)+[\bs r \to -\bs r]\Bigg\}+\nonumber\\+&\Bigg\{4\sin\Delta_{k,0}^se^{\pi\eta/2}\f R\Big[\c N\;\v F\big(1{+}i\eta,1,-i(kr{+}\bs k \bs r)\big) ie^{-i\Delta_{k,0}^s}e^{-2i\delta_{k,0}^c}U\big(1{-}i\eta,2,2ikr\big)\Big]+[\bs r \to -\bs r]\Bigg\}-\nonumber\\&-8\sin^2\Delta_{k,0}^se^{\pi\eta}\big|U\big(1{-}i\eta,2,2ikr\big)\big|^2 .
\end{align}
\end{widetext}

\subsection{The two-particle correlation function}
\label{ss:corrcalc}
In this section we combine the previously discussed approaches, and write up the complete functional form of the correlation function by plugging in Equation (\ref{eq:dcc}) and (\ref{eq:psiabs}) to Equation (\ref{eq:c2}).

\begin{align}
C_2(k) = 1-\lambda+\lambda\cdot\c I^{(c,c)}(k),
\end{align}
where the $\c I_{(c,c)}(k)$ integral can be written as
\begin{align}
\c I_{(c,c)}(k) &= \sint{-1mm}{-1mm}{}{} d^3\bs r D_{(c,c)}(\bs r)\big|\Psi^{(2)}_{\bs k}(\bs r)\big|^2 =\nonumber\\&= 2\pi\sint{-1mm}{-4mm}0\infty dr\;r^2D_{(c,c)}(r)\sint{-1mm}{-2mm}{-1}1dy\big|\Psi^{(2)}_{\bs k}(\bs r)\big|^2 .
\label{eq:icc}
\end{align}
Substituting Eq.~(\ref{eq:psiabs}) into Eq.~(\ref{eq:icc}) we get the following expression:
\begin{align}
&\c I^{(c,c)}(k) = 2\pi\Big\{|\c N|^2{\times}\c I^{(1)}(k) + |\c N|^2{\times}\c I^{(2)}(k)-\nonumber\\&-8\sin^2\Delta_{k,0}^s e^{\pi\eta}{\times}\c I^{(3)}(k) + 8\sin\Delta_{k,0}^s e^{\pi\eta/2}{\times}\nonumber\\&\times\f R\Big[i\c Ne^{-i\Delta_{k,0}^s}e^{-2i\delta_{k,0}^c}\c I^{(4)}(k)\Big]\Big\} ,
\label{eq:c2full}
\end{align}
where the following integrals were introduced:
\begin{alignat}{2}
&\c I^{(1)} {=} \sint{-1mm}{-4mm}0\infty\m dr\,r^2D_{(c,c)}(r)\sint{-1mm}{-3mm}{-1}1\m dy\,\big|\v F\big(1{-}i\eta,1,ikr(1{+}y)\big)\big|^2 ,\\
&\c I^{(2)} {=} \sint{-1mm}{-4mm}0\infty\m dr\,r^2D_{(c,c)}(r)\sint{-1mm}{-3mm}{-1}1\m dy\,\Big\{\v F\big(1{-}i\eta,1,ikr(1{+}y)\big)\times\nonumber\\&
\hspace{34mm}\times\v F\big(1{+}i\eta,1,-ikr(1{-}y)\big)\Big\},\\
&\c I^{(3)} {=} {2}\sint{-1mm}{-3mm}0\infty\m dr\,r^2D_{(c,c)}(r)\cdot\big|U\big(1{-}i\eta,2,2ikr\big)\big|^2,\\
&\c I^{(4)} {=} \sint{0mm}{-3mm}0\infty\m dr\,r^2D_{(c,c)}(r)\cdot U\big(1{-}i\eta,2,2ikr\big)\times\nonumber\\&
\hspace{30mm}{\times}\sint{-1mm}{-3mm}{-1}1\m dy\,\v F\big(1{+}i\eta,1,-ikr(1{+}y)\big).
\end{alignat}
%
The last step is to explore the dependence of the strong phase shift $\Delta_{k,0}^s$ on $k$. Using the notation of Ref.~\cite{Lednicky:2005tb} we can relate $\Delta_{k,0}^s$ to the full (Coulomb+strong) scattering amplitude $f_c(k)$:
\begin{align}
\sin\Delta_{k,0}^se^{i\Delta_{k,0}^s} = k|\c N|^2f_c(k).
\end{align}
The scattering amplitude $f_c(k)$ can be expressed as~\cite{LandauIII}
\begin{align}
f_c(k) = \bigg(\rec{K(k)}-2k\eta\Big(h(\eta)+i\frac{|\c N|^2}{2\eta}\Big)\bigg)^{-1},
\end{align}
where $h(\eta)$ is related to the digamma function $\psi$ as
\begin{align}
h(\eta) = \big[\psi(i\eta)+\psi(-i\eta)-\log(\eta^2)\big]/2.
\end{align}
The $k$ dependence of $f_c(k)$ partly comes from the function $K(k)$, which can be expressed with the $\delta_{k,0}^{(2)}$ phaseshift (where the $(2)$ superscript denotes the $I = 2$ isospin channel, the only allowed channel in case of identical charged pion pairs):
\begin{equation}
K(k) = \frac{1}{k}\tan\delta_{k,0}^{(2)}.
\end{equation}
If there would be no Coulomb, only strong interaction, $\delta_{k,0}^{(2)}$ would be identical to the previously introduced $\Delta_{k,0}^s$ strong phase-shift. One can find different parametrizations for $\delta_{k,0}^{(2)}$ in the literature, in the following we mention some of them. A simple parametrization can be found in J.~Bijnens~et~al.~\cite{Bijnens:1997vq}:
\begin{align}
\label{e:bijnens}
K(k) = \Bigg(\frac{m_\pi}{a_0^{(2)}}+\rec{2}r_0^{(2)}k^2\Bigg)^{-1},
\end{align}
where $a_0^{(2)}$ is called the scattering length, and $r_0^{(2)}$ is called the effective range. The latter can also be connected to a $b_0^{(2)}$ slope parameter as
\begin{align}
r_0^{(2)} = \rec{m_\pi a_0^{(2)}} -\frac{2m_\pi b_0^{(2)}}{\big(a_0^{(2)}\big)^2}-\frac{2a_0^{(2)}}{m_\pi} .
\end{align}

This effective-range parametrization is thought to be useful when the scattering length is much larger than the range of the scattering potential~\cite{Adhikari:2018ukk}, which is not the case for identical pion scattering. Another parametrization~\cite{Schenk:1991xe} better suited for our investigations can be written up with the help of the center-of-mass energy $s~=~4(m_\pi^2+k^2)$ as

\begin{align}
K(k) = \frac{2}{\sqrt{s}}\frac{4m_\pi^2{-}s_0^{(2)}}{s{-}s_0^{(2)}}\Bigg(a_0^{(2)}{+}\tilde{b}_0^{(2)}\frac{k^2}{m_\pi^2}\Bigg),\textnormal{ where}
\label{e:cgl}
\end{align}
\begin{align}
\tilde{b}_0^{(2)} = b_0^{(2)}{-}\frac{4m_\pi^2 a_0^{(2)}}{s_0^{(2)}-4m_\pi^2}.
\end{align}

The $s_0^{(2)}$ parameter corresponds to the value of $s$ where the phase shift passes through 90$^\circ$. It usually has a negative value, indicating that for the $I = 2$ channel the phase remains below 90$^\circ$. The parametrization can also be extended with higher order terms, the values of the parameters can be found e.g. in Colangelo-Gasser-Leutwyler (CGL)~\cite{Colangelo:2001df}: $a_0^{(2)} = -0.0444,\; b_0^{(2)} = -0.0803\;m_\pi^{-2},\; s_0^{(2)} = -21.62\;m_\pi^2.$

\begin{figure}
\centerline{\includegraphics[width=0.5\textwidth]{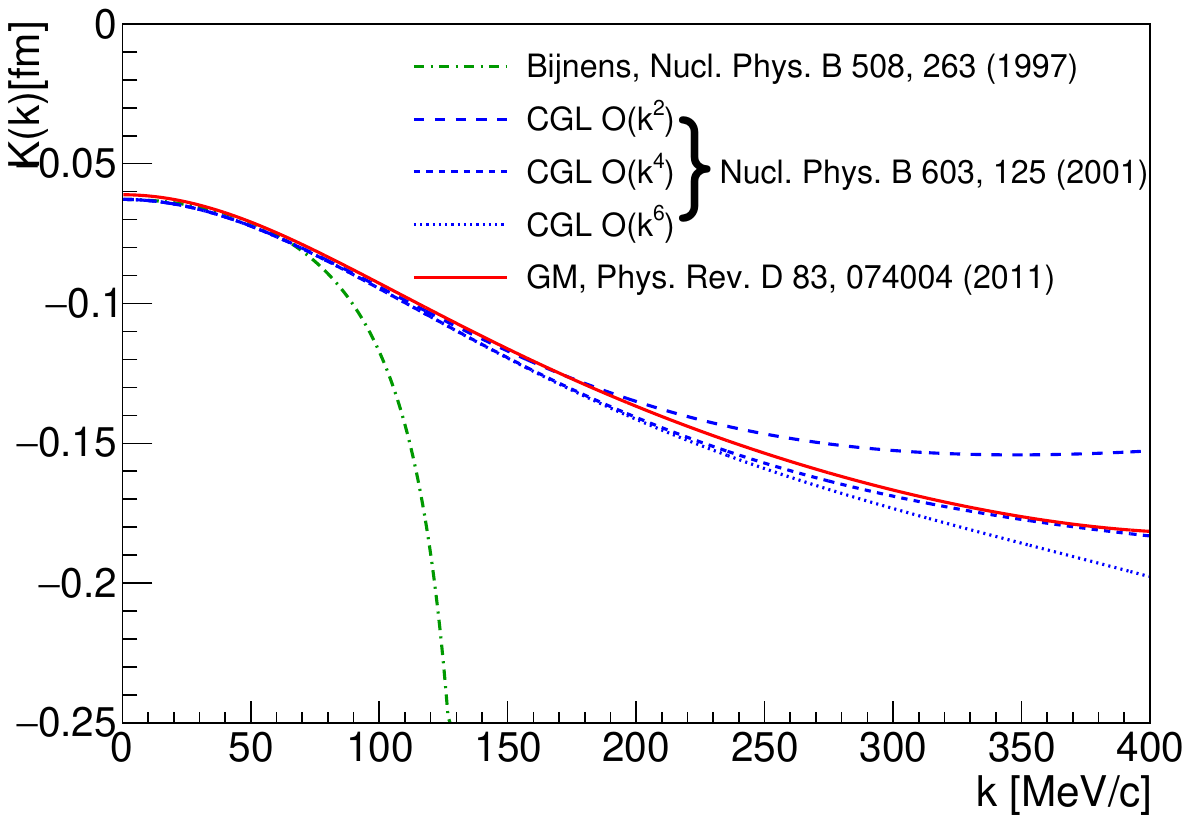}}
\caption{Comparison of different $K(k)$ parametrizations. See equations (\ref{e:bijnens}), (\ref{e:cgl}), and (\ref{e:gm}) for Bijnens, CGL and GM, respectively.}
\label{f:Kpar}
\end{figure}

A different parametrization can be found in a more recent paper from Garc\'ia-Mart\'in~et~al. (GM)~\cite{GarciaMartin:2011cn}:
\begin{align}
K(s) = \frac{2}{\sqrt{s}}\frac{s-2z_2^2}{m_\pi^2}\Bigg(B_0{+}B_1\frac{\sqrt{s}-\sqrt{\hat{s}-s}}{\sqrt{s}+\sqrt{\hat{s}+s}}\Bigg)^{-1},
\label{e:gm}
\end{align}
where the parameter values are the following: $z_2 = 143.5 \textnormal{ MeV}, B_0 = -79.4, B_1 = -63.0, \sqrt{\hat{s}} = 1050 \textnormal{ MeV}$. A comparison of the previously mentioned parametrizations can be seen on Fig.\ref{f:Kpar}. In the $k$ range important for our investigations ($k\lesssim100$ MeV/c) the different parametrizations give almost identical results, so in the following we utilized the most recent one from Ref.~\cite{GarciaMartin:2011cn}.

\section{Numerical results}
\label{s:results}
In this chapter we present the results of the numerical calculation of $C_2(k)$. Using numerical integral calculations we created a lookup table for the function defined in Equation (\ref{eq:c2full}) for a wide range of values of $k$, $R$ and $\alpha$. This lookup table then was used to obtain the value of the function for any $k, R$ and $\alpha$ by interpolation (within the available range).

If we omit the $\c I^{(3)}$ and $\c I^{(4)}$ terms from Eq. (\ref{eq:c2full}), we get back the pure Coulomb part. In the following, we compare the correlation function containing only the Coulomb interaction with the one containing both the Coulomb and the strong interactions, and try to give an estimate on the change in the values of the L\'evy source parameters that is caused by the proper treatment of the strong interaction compared to the neglection of it.

From here on, we change the relative momentum variable to $Q = 2k$ to better compare to the notation of published experimental results. 

\subsection{Comparison of Coulomb and strong FSI effects}
\label{ss:strongfsi}
\begin{figure}[t]
\centerline{\includegraphics[width=0.48\textwidth]{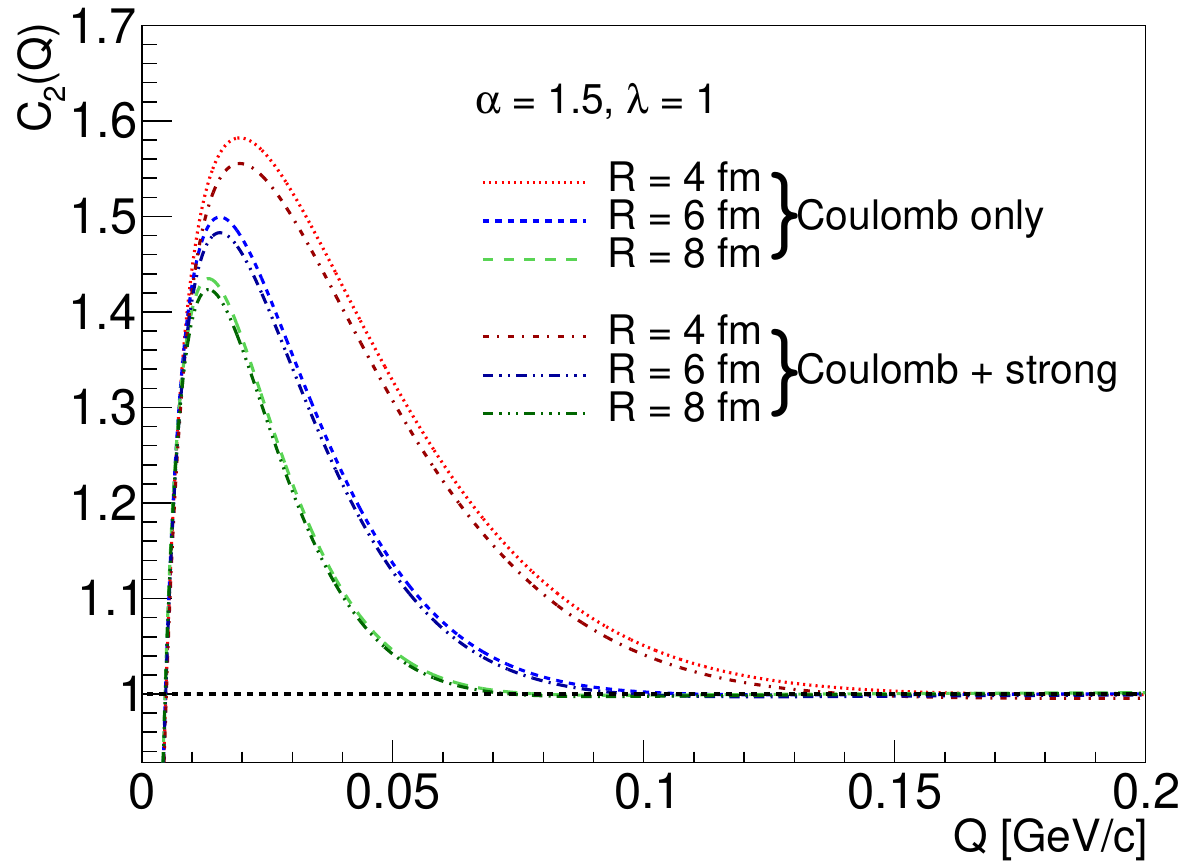}}
\caption{Two-pion correlation functions calculated for L\'evy-stable sources. Three different L\'evy-scale values are compared at the same index of stability $\alpha = 1.5$ and same correlation strength $\lambda = 1$. The functions containing only the Coulomb interaction and the ones including both the Coulomb and strong interactions are shown separately.}
\label{f:c2examples}
\end{figure}

Fig.~\ref{f:c2examples}. shows the calculated correlation functions for three different L\'evy-scale values at the same index of stability $\alpha$ and same correlation strength $\lambda$. It is clearly visible that turning on the strong interaction affects the strength of the correlation functions, however, the effect on the L\'evy-scale $R$ and the index of stability $\alpha$ is not so transparent at this point. 

\begin{figure}[t]
\includegraphics[width=0.48\textwidth]{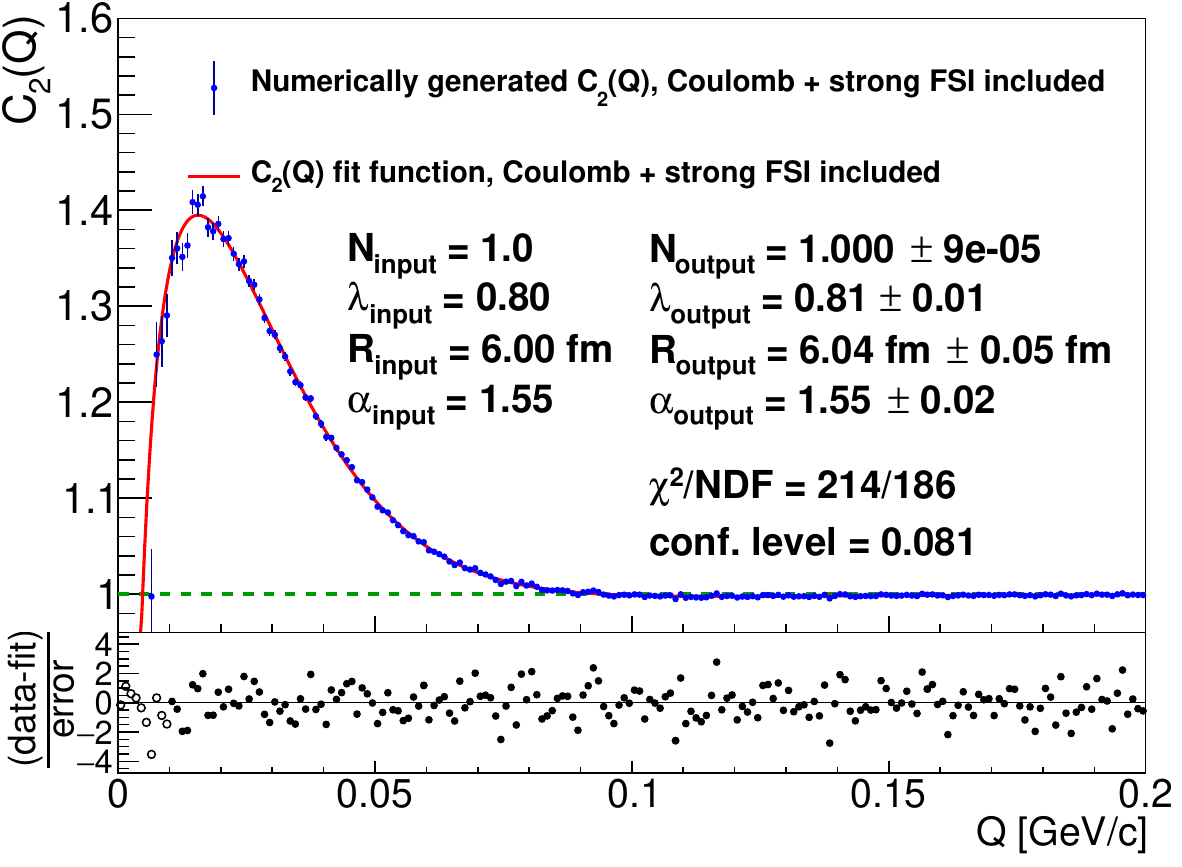}
\caption{Numerically generated two-pion correlation histogram, fitted with the corresponding functional form to test the validity of the fitting method. The output parameter values are within errors the same as the input.}
\label{f:strongdata_strongfit}
\end{figure}

To investigate the effect of the strong interaction in more detail, we generated histograms by sampling the calculated functions containing both Coulomb and strong interactions. To make the generated correlation function resemble real data, we randomly scatter the points around the calculated function and assign a relative error proportional to $1/Q$ (which is a realistic assumption if one considers typical experimental scenarios). We then fit the generated data with the help of the ROOT Minuit2 minimizer framework, with a similar method to what is described in Ref.~\cite{Adare:2017vig}. To check the validity of the fitting method, first we fit the generated histogram with the corresponding functional form to see if we get back the input parameter values. Fig.~\ref{f:strongdata_strongfit} shows such a fit to the generated data. The fit converged with an acceptable $\chi^2$/NDF value, the error matrix turned out to be accurate, and for the output parameter we got back within errors the same ones as were given as input. We repeated this test for multiple different input parameter values and found that our fitting method is indeed reliable.

\begin{figure*}[ht]
\includegraphics[width=0.48\textwidth]{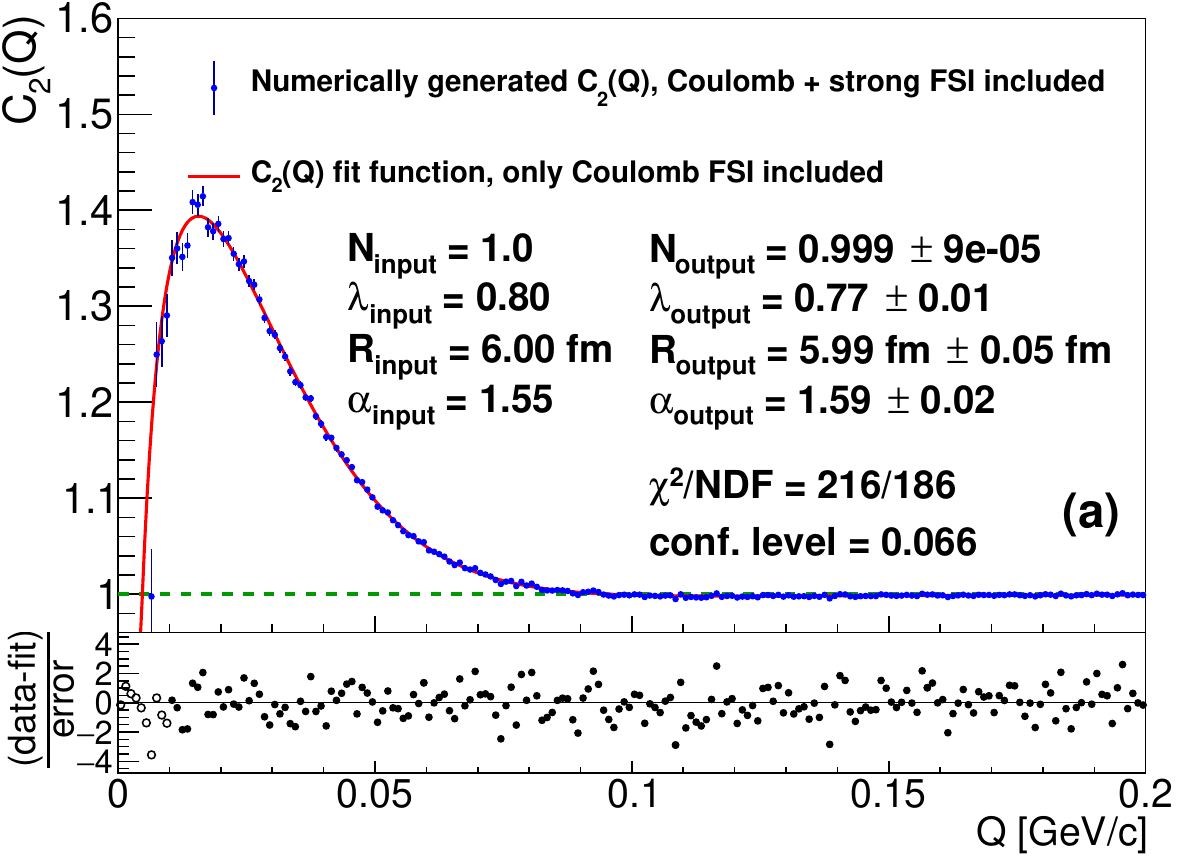}
\includegraphics[width=0.48\textwidth]{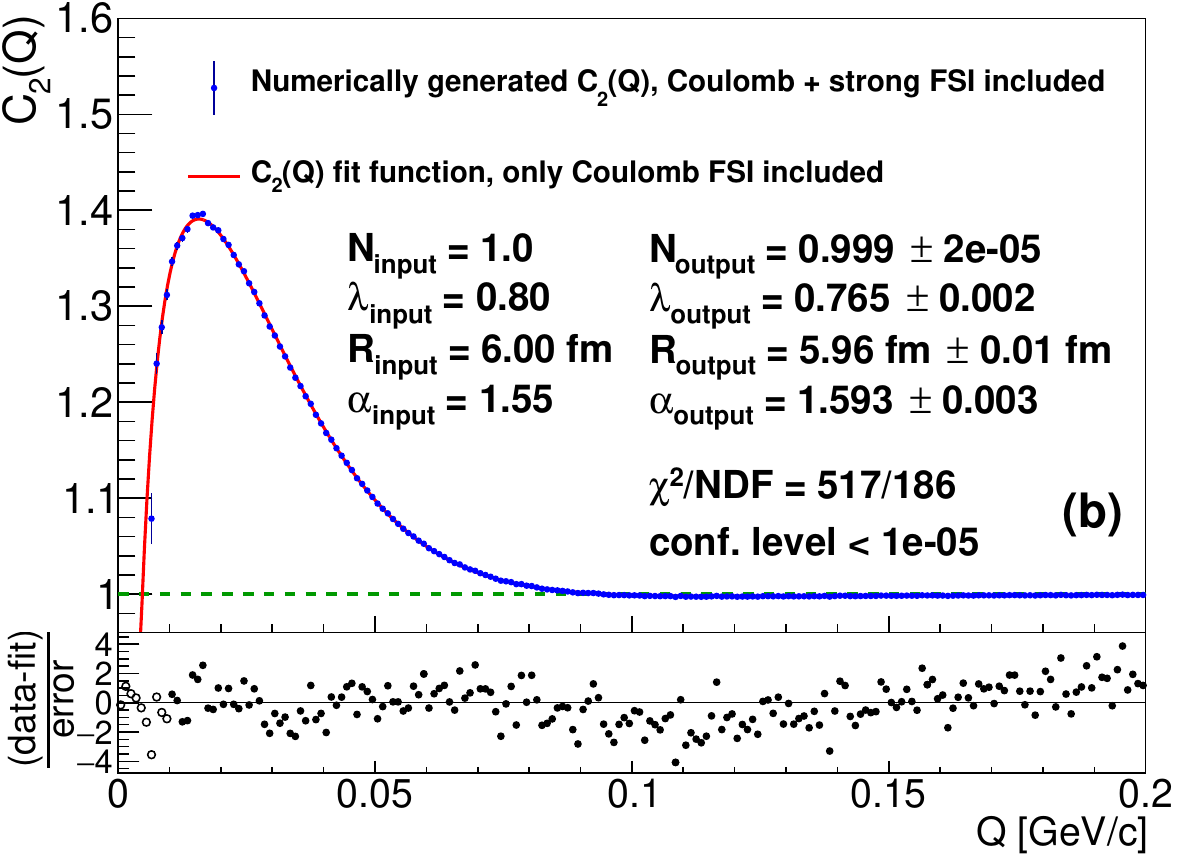}
\caption{Numerically generated two-pion correlation histogram incorporating Coulomb and strong final state interactions, fitted with a functional form containing only the Coulomb effect. When the generated data is less precise (a), the fit is statistically acceptable, but the output parameter values differ from the input. The difference is even more pronounced when the generated data is more precise (b), in this example the value of $\lambda$ decreased by about 4\%, the value of $R$ decreased by about 1\%, and the value of $\alpha$ increased by about 3\%. It is also important to note that in this case the $\chi^2/$NDF value is not acceptable anymore.}
\label{f:strongdata_coulombfit}
\end{figure*}

As a next step, we took the same generated data and fitted it with a function containing only the effect of the Coulomb interaction. Fig.~\ref{f:strongdata_coulombfit} shows an example for such a fit on panel (a). The fit converged again, the error matrix again turned out to be accurate. The resulting $\chi^2$ value becomes just slightly higher than before, nevertheless, the fit is still acceptable. Although in this case the function containing only the Coulomb interaction can provide an acceptable fit to the generated data which contains also the strong interaction, the values of the fit parameters differ from the input parameter values. It seems that in this case one underestimates the value of $\lambda$ from such a fit, and overestimates $\alpha$. Within this precision, it seems that the value of $R$ is unaffected.

One can also assume that if the data is more precise, meaning that the fluctuation and the statistical uncertainty of the generated points are smaller, the fit will not provide an acceptable $\chi^2$ anymore. To check this, we also generated such $C_2(Q)$ histrograms, and found that the Coulomb fits converged, but indeed the $\chi^2$ values increase by a considerable amount resulting in statistically unacceptable fits. An example for this can be seen on panel (b) of Fig.~\ref{f:strongdata_coulombfit}. One can also observe that on the subplot showing the values of the difference of the fit from the data divided by the uncertainty of the datapoint, a characteristic oscillating structure appears.

\subsection{Quantitative estimation of the strong FSI effect}

\begin{figure*}[ht]
\includegraphics[width=0.325\textwidth]{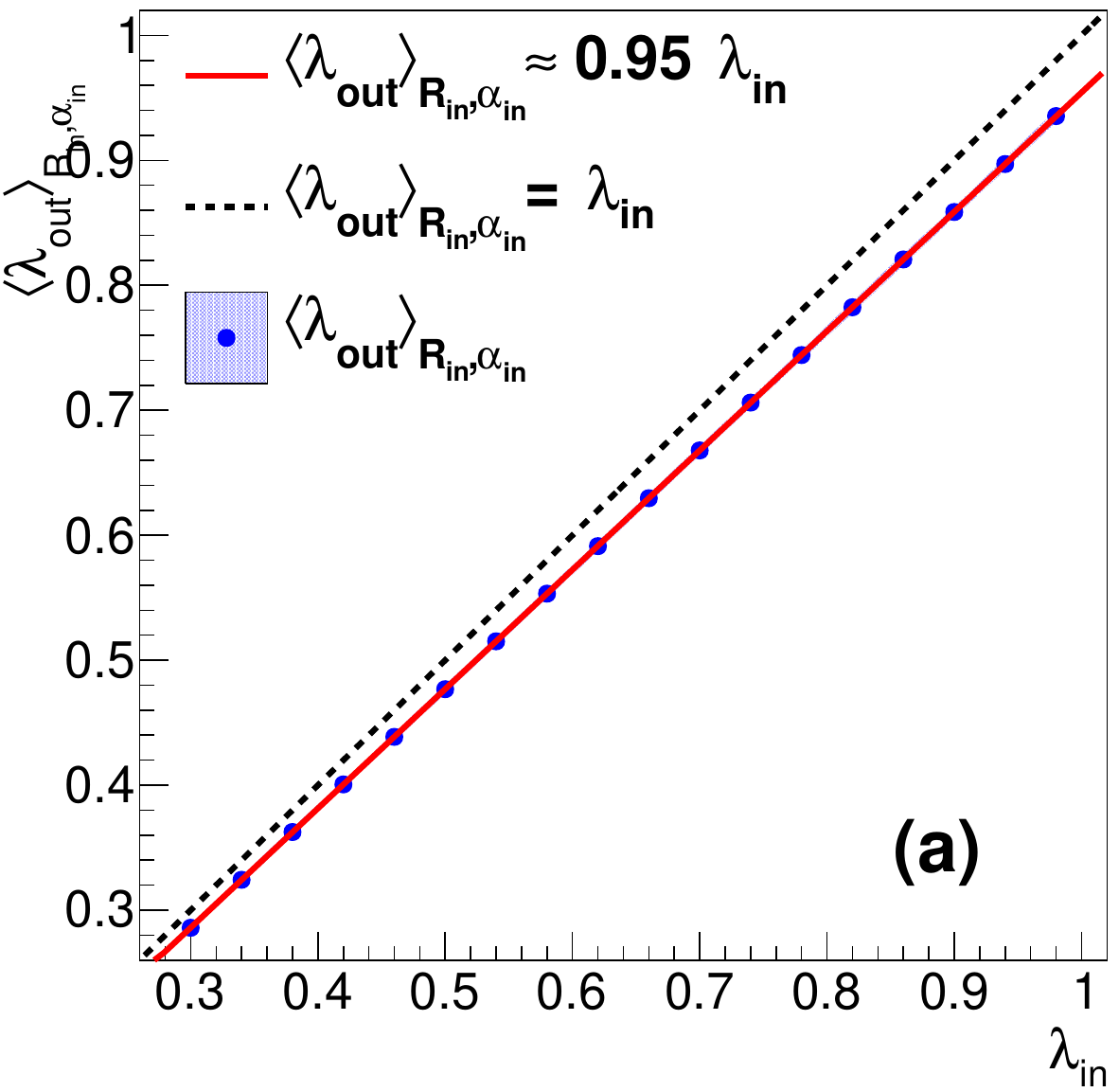}
\includegraphics[width=0.325\textwidth]{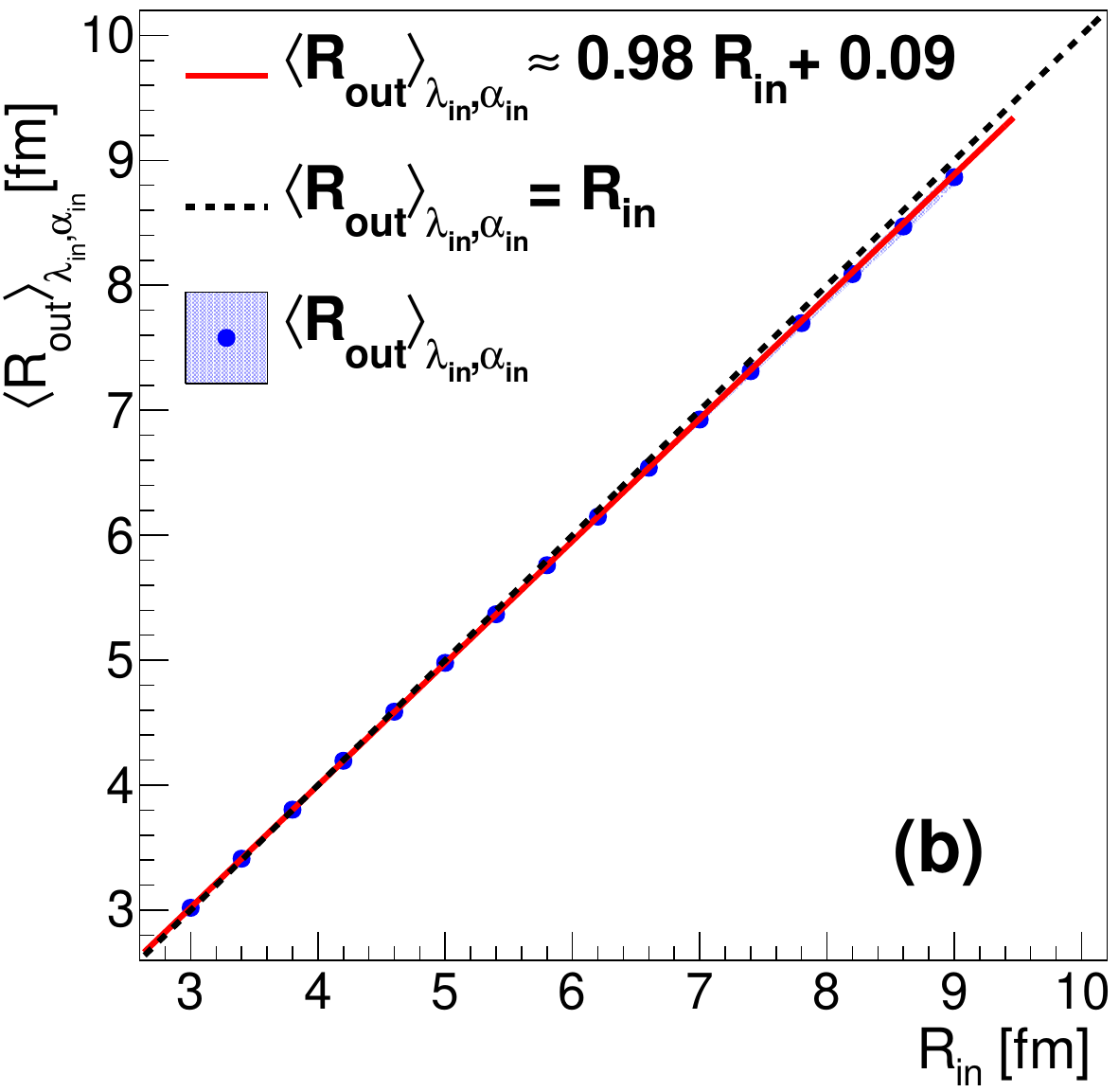}
\includegraphics[width=0.325\textwidth]{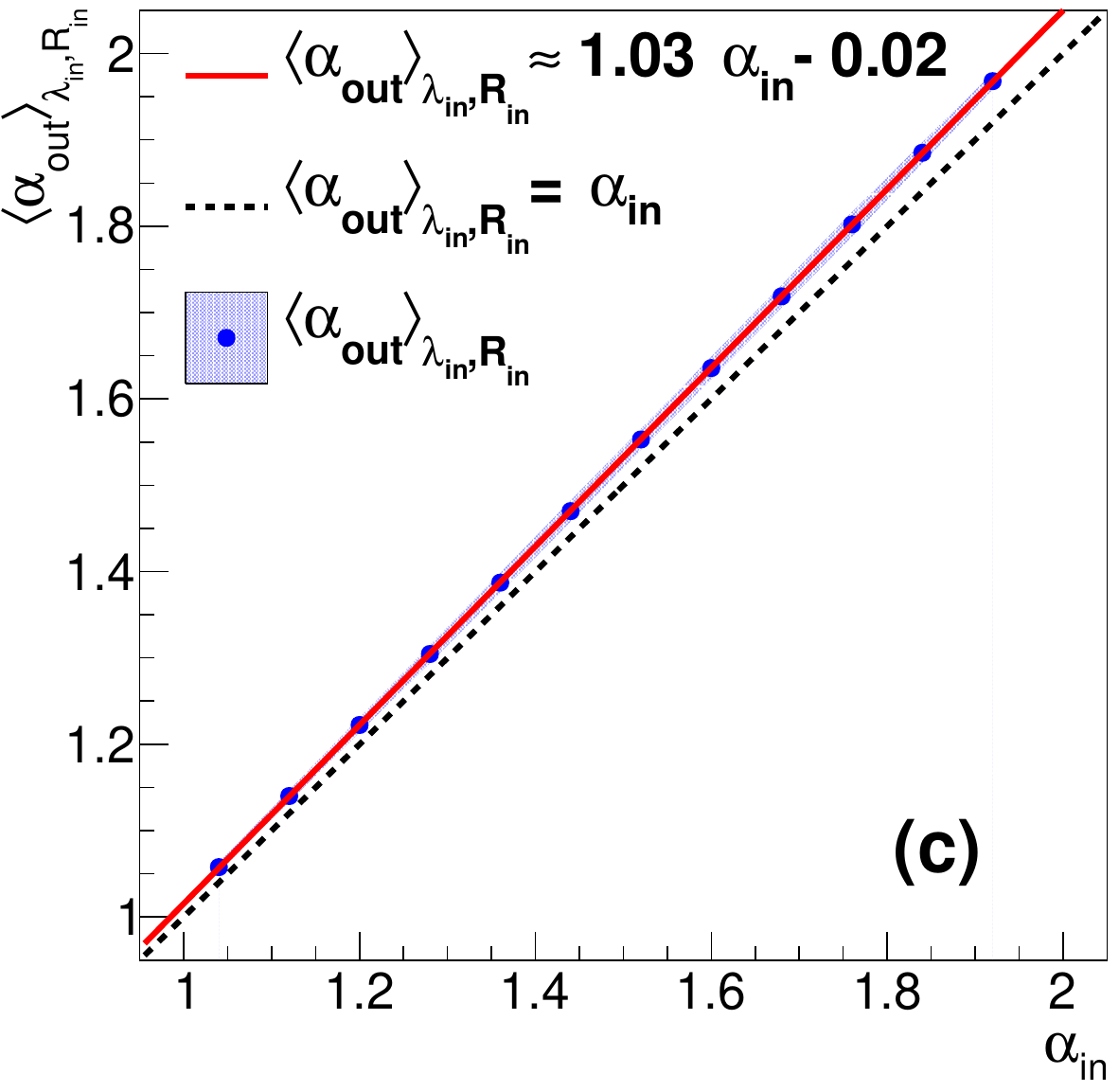}
\caption{Output versus input values from fits similar to Fig.~\ref{f:strongdata_coulombfit}.(b). The correlation strength $\lambda$ is shown on panel (a), the L\'evy scale parameter $R$ is shown on panel (b) and the L\'evy exponent $\alpha$ is shown on panel (c). The identity line is shown with a dashed line, while a linear fit is shown with a continuous line. For a given input parameter, the weighted average of the output values are shown with markers, and the standard deviation is shown with a band.}
\label{f:inout}
\end{figure*}

To give a better estimation on the change in the parameter values when fitting data containing strong interaction with a function containing only the Coulomb effect, we generated and fitted histograms similar to panel (b) of Fig.~\ref{f:strongdata_coulombfit}, spanning a wide range in parameter space of $\lambda_{input} = 0.3-1.0$, $R_{input}$ = 3 fm - 9 fm and $\alpha_{input} = 1.0 - 2.0$. For each fit parameter, we plotted the output versus the input values. The plotted output values represent a weighted average of output values coming from the same input for the given parameter but different inputs for the other two parameters. The results of this investigation can be seen on Fig.~\ref{f:inout}, panel (a)--(c).

By fitting data containing the Coulomb and strong final state interactions with a functional form describing only the Coulomb part, it seems that the correlation strength $\lambda$ is underestimated by about 5\% on average. The effect on the L\'evy-scale parameter $R$ is negligible at small values of it, while at higher values of $R$ (up to about 9 fm) it is also slightly underestimated, by about 1\%. The L\'evy exponent $\alpha$ is overestimated by about 1-2\%.

The estimations given here for the change in parameter values are by no means universal, they also depend on other factors such as numerical precision of the integral calculations, fit limits ($Q_{min}$ dependence), the precision of the generated data (see for example the difference between Fig.~\ref{f:strongdata_coulombfit} (a) and (b)), or the parametrization of the strong phase-shift. The important conclusion from our investigations is that if the data is precise enough (which could be the case for recent measurements at RHIC or LHC), one most likely has to incorporate the strong interaction in the fits to achieve a statistically acceptable description of pion-pion correlation functions.



\section{Summary and conclusions}

In this paper we presented a detailed calculation of the shape of two-pion HBT correlation functions with the assumption of L\'evy stable source functions, and taking into account the Coulomb and strong final state interactions. Strong final state interactions were treated in the s-wave approximation.

A numerical calculation of the correlation function revealed that the strong final state interaction can have a non-negligible effect on the shape of pion-pion correlation function. As a first step towards the more thorough evaluation, we presented a quantitative estimation of the magnitude of this effect. As a general trend, we can ascertain that fits without the strong interaction effect typically underestimate the strength of the correlation, $\lambda$, and the L\'evy scale $R$, while overestimate the L\'evy exponent $\alpha$. The magnitudes of these deviations are generally found to be no more than a few percent.

However, typical fits to measured correlation functions can become statistically unacceptable if the strong interaction is neglected. If one aims at a high level of precision (feasible in case of precise enough data coming from today's typical heavy ion experiments), one can arrive at refined conclusions about the source function if the small deviations (caused by the strong interaction) are treated properly in the fitting procedure. 

As an outlook, we note that there is some room for improvement in the methodology of the numerical calculations presented here. Such improvements might yield so precise predictions that it becomes possible to actually give constraints on like-sign pion strong interactions (i.e. scattering lengths) based on HBT correlation measurements in heavy ion collisions, a topic long thought to be interesting to investigate~\cite{Sumbera:2007ir}. We look forward to a concrete experimental test of the predictions made here about the shape of the correlation function that gets influenced by strong final state interaction.

\label{s:summary}

\section*{Acknowledgments}
The authors would like to thank R. Lednick\'y, D. Mi\'skowiec and T. Cs\"org\H{o} for useful discussions. Our research has been partially supported by the Hungarian NKIFH grants No. FK-123842 and FK-123959. The authors were also supported by the \'UNKP-19-3 and \'UNKP-19-4 New National Excellence Program of the Hungarian Ministry for Innovation and Technology. M. Csan\'ad and M. Nagy were supported by the J\'anos Bolyai Research Scholarship. 

\appendix

\section{Assorted special functions}
\label{s:app:formulas}
The following definitions, formulas and the explanation of the special functions that come by can be found in any standard textbook on quantum mechanical scattering theory (such as Ref.~\cite{LandauIII}), nevertheless we write them up to make the paper as self-contained as possible.

In the treatment of the quantum mechanical Coulomb problem, one encounters the {\it confluent hypergeometric equation}, a second order linear differential equation for the unknown $f(z)$ function, written as 
\begin{align}
zf''(z) + (b{-}z)f'(z) -af(z) = 0,
\end{align}
where $a$ and $b$ are two arbitrary parameters. A commonly used pair of linearly independent solutions are provided by the (renormalized) {\it confluent hypergeometric function} or Kummer's function:
\begin{align}
&\v F(a,b,z) := \frac{F(a,b,z)}{\Gamma(b)},\\
&F(a,b,z) := \sum_{n=0}^\infty\frac{\Gamma(a{+}n)\Gamma(b)}{\Gamma(a)\Gamma(b{+}n)}\frac{z^n}{n!},
\end{align}
which has the convenient property that it is analytic everywhere, especially at $z{=}0$; and the other solution is the so-called Tricomi's function, defined as 
\begin{align}
U(a,b,z) = &\frac{\pi}{\sin(\pi b)}\bigg\{\frac{\v F(a,b,z)}{\Gamma(a{+}1{-}b)}-\nonumber\\ &\qquad-z^{1-b}\frac{\v F(a{+}1{-}b,2{-}b,z)}{\Gamma(a)}\bigg\}
\end{align}
if $b$ is not an integer, and as a limit $b\to n$ in the $b=n{\in}\mathbb Z$ integer case. The $U(a,b,z)$ function has a branch point at $z{=}0$, with the form written up having a branch cut along the $z{\in}\mathbb R^-$ negative real line. However, it has the convenient property that it behaves asymptotically as
\begin{align}
U(a,b,z)\sim z^{-a},
\end{align}
and this is a property that is unique to it among the solutions of the confluent hypergeometric equation.

A ,,dual'' pair of useful properties of the functions introduced is
\begin{align}
\v F(a,b,z) = e^z\v F(b{-}a,b,-z),\\
U(a,b,z) = z^{1-b}U(a{+}1{-}b,2{-}b,z),
\end{align}
the former of which is verified by noting that both sides are analytic and fulfill the very same differential equation; the latter is a simple consequence of the definition. As seen above, $U(a,b,z)$ can be expressed from $\v F(a,b,z)$; one can also derive the ,,inverse'' formula:
\begin{align}
\v F(a,b,z) = &\frac{e^{i\Pi_za}}{\Gamma(b{-}a)}U(a,b,z) +\nonumber\\ &\qquad+\frac{e^{i\Pi_z(a-b)}}{\Gamma(a)}e^zU(b{-}a,b,{-}z),\label{e:FfromU}
\end{align}
with the $\Pi_z$ notation introduced here as being $\pi$ or $-\pi$, if $\operatorname{arg}z{>}0$ or $\operatorname{arg}z{<}0$, respectively.

Using l'Hospital's rule, the power series expression of the $U(a,b,z)$ function for integer $b$ turns out to be
\begin{align}
&U(a,m{+}1,z) =\frac{(-1)^m}{\Gamma(a{-}m)}\bigg\{-\log z\cdot\v F(a,m{+}1,z) +\nonumber\\&{+} \sum_{s=1}^m\frac{(-1)^s}{z^s}\frac{(s{-}1)!}{(m{-}s)!}\frac{\Gamma(a{-}s)}{\Gamma(a)}{+} \sum_{s=0}^\infty\frac{z^s}{s!}\rec{(m+s)!}\frac{\Gamma(a{+}s)}{\Gamma(a)}\times\nonumber\\&{\times}\big[\psi(s{+}1){-}\psi(a{+}s){+}\psi(s{+}m{+}1)\big]\bigg\},\qquad m{\in}\B N_0^+.
\end{align}
Here $\psi(s)$ is the digamma function defined as
\begin{align}
\psi(s) \equiv \frac{\Gamma'(s)}{\Gamma(s)}.
\end{align}
Some convenient properties of it are:
\begin{align}
\psi(a{+}n) &= \psi(a) + \sum_{k=1}^n\rec{a{+}k},\\
\bfollows\quad
\psi(n{+}1) &= -\gamma + \sum_{k=1}^n\rec k,
\end{align}
where $\gamma$ is the Euler constant:
\begin{align}
\gamma = \lim_{n\to\infty}\bigg(\sum_{k=1}^n\rec k-\ln n\bigg) = 0.577\dots
\end{align}
A side note to the calculation of the $\v F(a,b,z)$ and $U(a,b,z)$ functions: for the typical parameter values encountered in our work (i.e. $a$ and $b$ on the order of unity), the power series in $z$ can be used in a numerically satisfactory way only up to $|z|\approx30$. For higher $|z|$ values, one rather uses the asymptotic expansion of $U(a,b,z)$:
\begin{align}
&U(a,a{+}1{-}\beta,z) = z^{-a}\bigg\{1 -\frac{a\beta}{1!z} +\frac{a(a{+}1)\,\beta(\beta{+}1)}{2!z^2}-\nonumber\\ &\quad\qquad-
\frac{a(a{+}1)(a{+}2)\,\beta(\beta{+}1)(\beta{+}2)}{3!z^3}+\ldots\bigg\},
\end{align}
and for $\v F(a,b,z)$, the expression of it that uses $U(a,b,z)$, see Eq.~(\ref{e:FfromU}) above.

Regrettably, most numerical packages that are used for the computation of special functions do not have built-in methods for the calculation of the gamma function and the digamma function, $\Gamma(z)$ and $\psi(z)$ for arbitrary complex arguments, which was very much needed for our objectives for this work. In our calculations, we used the Lanczos approximation~\cite{Lanczos:1964zz} for both $\Gamma(z)$ and $\psi(z)$ when it was necessary. Usually, the Lanczos approximation is written up only for $\Gamma(z)$, however, it is easy to verify that the approximative formula is a well-behaved smooth function of $z$, so it can safely be used for the calculation of $\psi(z)$ as well, by taking the logarithmic derivative of it.

\end{document}